**Decentralized physiology and the molecular basis of social life in eusocial insects**

DA Friedman [1*], BR Johnson [1‡], TA Linksvayer [2‡].

1. University of California, Davis, Department of Entomology. Davis, CA, 95616.
2. University of Pennsylvania, Department of Biology. Pennsylvania, PA, 19104.

* Corresponding Author. Email: DanielAriFriedman@gmail.com , Address: One Shields Avenue, 383 Briggs Hall, Davis, CA, 95616.
‡ These authors contributed equally to the work.

**Abstract.**

The traditional focus of physiological and functional genomic research is on molecular processes that play out within a single body. In contrast, when social interactions occur, molecular and behavioral responses in interacting individuals can lead to physiological processes that are distributed across multiple individuals. In eusocial insect colonies, such multi-body processes are tightly integrated, involving social communication mechanisms that regulate the physiology of colony members. As a result, conserved physiological mechanisms, for example related to pheromone detection and neural signaling pathways, are deployed in novel contexts and regulate emergent colony traits during the evolutionary origin and elaboration of social complexity. Here we review conceptual frameworks for organismal and colony physiology, and highlight functional genomic, physiological, and behavioral research exploring how colony-level traits arise from physical and chemical interactions among nestmates. We highlight mechanistic work exploring how colony traits arise from physical and chemical interactions among physiologically-specialized nestmates of various developmental stages. We consider similarities and differences between organismal and colony physiology, and make specific predictions based on a decentralized perspective on the function and evolution of colony traits. Integrated models of colony physiological function will be useful to address fundamental questions related to the evolution and ecology of collective behavior in natural systems.

**Colony Organization = Social Anatomy + Social Physiology**

Eusocial colonial insects, such as ants, termites, and honey bees, thrive across almost all terrestrial ecosystems [1,2]. The ecological success of these species rests in their use of colony traits, such as nest architecture [3,4] and collective foraging behavior [5–7], which are functionally absent in solitary insects yet keenly developed in eusocial taxa. In the eusocial insects, division of labor (DOL) describes how nestmates vary in their form and function. DOL formalizes the extent of specialization among nestmates in the performance of tasks, usually with a physiological or morphological basis or arising from nestmate age (temporal polyethism) and experience [8–11]. We build off of the conceptual framework of Johnson & Linksvayer [12] that considers eusocial colony organization from the perspective of **social anatomy** & **social physiology**:

**Social anatomy** is the notion that colonies are composed of specialized parts with limited roles, like the organs of an individual animal. Specialized colony anatomy allows for greater productivity and efficiency for the completion of many tasks. Physiological specialization exists at multiple levels within the colony. Queen-worker specialization allows for an increased reproductive output of queens alongside increased work output from workers, from the same diploid female genome [13]. Subspecialization among workers can manifest as permanent variation in body morphology [14], temporal polyethism, the process by which stereotyped changes in worker tissue-specific physiology (via changes in gene expression [15,16]) lead to differential reactivity to stimuli related to various inside and outside tasks, and other forms of specialization between workers [17]. The primary anatomical division within the colony, like a multicellular organism, is soma-germline, e.g. between reproductive and non-reproductive components of the colony. Just as there is extreme variation in body plan and life history across multicellular organisms, there is variation among eusocial taxa in a number of colony parameters, such as colony size, queen number, nest architecture, and reproductive life history [18]. Though the soma-germline distinction is blurry in some eusocial lineages (e.g. in the ant taxa of *Harpegnathus* and *Ooceraea* where workers retain some reproductive potential) the same can be said of many multicellular organisms (plants, worms, sponges) [19–21].

**Social physiology** is the set of dynamic mechanisms that coordinate the activity and development of the specialized parts of a colony. The principles of colony physiology are broadly the same as organismal physiology (e.g. homeostasis, hormesis, balance of anabolism/catabolism, nutrient partitioning among tissues), as colonies are complex adaptive systems that are targets of selection (successful colonies leave more offspring colonies, just as successful organisms leave more offspring organisms). Colony physiological mechanisms transfer information among nestmates and include both physical interactions (vibrations and tactile contact) and chemical signaling [12,22,23]. Here we consider a range of chemical signaling mechanisms involved in colony physiology: volatile and nonvolatile pheromones [23–25], as well as the direct transfer of bioactive compounds such as small RNAs, proteins, hormones, and nutrients [26–28]. Members of the eusocial insect colony implement these physiological mechanisms through their own body processes (e.g. the kinds of mechanisms homologous to solitary insects), as well as by playing roles in larger and slower colony-level physiological processes such as the regulation of development or foraging behavior. This is analogous to consideration of rapid and short-scale intracellular physiology, alongside the role of a given cell type within slower multi-organ physiological pathways.

The eusocial insects utilize many of the key molecular players that regulate behavior in solitary insects, such as the biogenic amine neurotransmitters and circulating hormones [29,30]. Specific hormones identified in solitary insects, such as corazonin and juvenile hormone are known to play important roles in the regulation of colony outcomes in eusocial insects [27,31–33]. Functional genomics has revealed that underlying the use of these conserved signaling pathways there is considerable molecular evolution: changes in protein-coding sequences of homologous genes, and larger-scale changes in gene family content [16,34,35]. Genome evolution and turnover happen at multiple scales, for example when compared to other Hymenoptera, ants display expansions of gene families related to odorant perception and other functional classes [34,36,37]. Gene family expansion and positive selection on odorant receptors may reflect lineage-specific selection on colony behavior [38,39], as well as shifts in ecological niche, for example in corbiculate bees [40]. It is unknown which components of odorant receptor evolution in eusocial taxa are related specifically to changes in communication amongst nestmates versus changes in the ecology of the species [1,23,38]. And while it is important to understand the receptivity of nestmates to various chemical cues, emphasis on the evolution of the primary sensory organs (e.g. chemoreceptor affinities) results in a reduced understanding of how intra-worker physiology differs in eusocial insects as compared to their solitary ancestors. Here we extend work on the evolution of social insect physiology beyond consideration of nestmate communication strategies [23], and towards a unified model of colony physiological function and evolution.

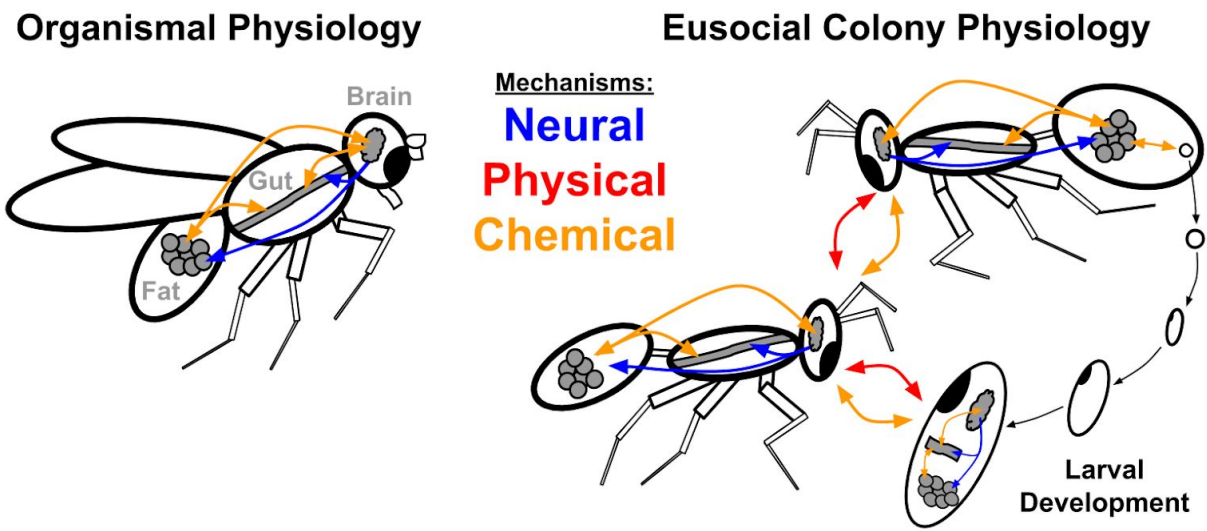

**Figure 1. Organismal and Eusocial Colony Physiology.** The regulation of nestmate variation and colony traits arises from interactions within and across the bodies of multiple castes, task groups, and developmental stages.

**Similarities between Colony and Organismal Physiology**

There are many functional and evolutionary analogies between eusocial insect colonies and multicellular organisms [41]. One approach that has often been discussed is to consider the eusocial insect colony as an "organism" [42] or a "superorganism" [43]. Whether or not one uses the "super-" to describe the eusocial insect colonial organism, enough analogies exist between colonies and multicellular organisms to warrant a functionalist approach. In colonies and organism bodies, similar principles are at play, such as decentralized transport, modularity [11,44], and metabolic scaling [45,46]. Hence, there is practical utility to taking an organismal approach (e.g. a functional perspective [47]) to study the eusocial colony function. The genetic consequences of eusociality are explored elsewhere [48–50]. Here we focus on the functional (i.e. physiological, molecular, and neural) and evolutionary implications of complex social life.

Colony and organismal physiology are both dynamic processes that play out via regulatory interactions across different tissues (Figure 1). The colony reflects a special higher-order structure where physiological subunits (nestmates) are integrated into a larger functional whole. In organismal physiology, we see multi-tissue neurohormonal pathways, for example in *Drosophila* "neural signals in the brain –> endocrine production in gut and fat cells –> alterations in foraging behavior & fat cell metabolism" [51,52]. In the eusocial insects, the coordination of foraging behavior with fat metabolism is also regulated by multi-tissue feedback loops (e.g. larval organs –> larval behavior –> worker brain –> worker organs). The hormonal and neurobiological mechanisms involved in the regulation of foraging in eusocial insects are conserved from similar systems in solitary insects [29,30,53]. Relative to how such mechanisms function in solitary insects, in eusocial insects these ancestral mechanisms become embedded within additional levels of colony-level regulation [54,55], facilitating the evolution of physiological specialization and decentralization.

In multicellular organisms just as in colonies, these multi-tissue physiological mechanisms are mediated by diffusible signaling molecules. In organismal physiology, conditioned fluids carry diffusible factors. These fluids include hemolymph (insects), as well as blood, lymph, and other fluids (mammals). In the eusocial insect colony, there is sharing of diffusible signaling molecules through the air (volatile compounds), through liquid solvents (through trophallaxis), as well as via the solid phase (deposition of long-lasting pheromone compounds on the ground allowing stigmergy). One key difference is that for physiological mechanisms that occur at the colony level, the multiple tissues involved are sometimes across multiple insect bodies – for example the fat body and brain of the larvae, as well as the brain and exocrine glands of the nurse. The decentralization of physiological processes across multiple nestmate bodies reflects the changes in colony function as contrasted with solitary insect biology.

Physiological mechanisms occur within single cells via signaling molecules [56], within insect bodies via endocrine signaling [57,58], and among insect bodies via exocrine signaling [23,25]. More than 75 distinct exocrine glands are known to exist in ants [43,59–61]. Many of these glands are known to secrete factors that regulate the behavior of nestmates, while most glands are of unknown specific function. In the case of eusociality, these exocrine secretory mechanisms have become embedded within colony-level decentralized physiological mechanisms: they are playing a fundamentally endocrine (internal regulatory) role within the colony. Whether one considers colony pheromones as exocrine compounds (from the perspective of the insect body glandular structure) or as colony endocrine compounds (from the perspective of the colony as an organism), there are key similarities between the influence of pheromones on workers and hormones on organs. Both colony pheromones and organismal hormones result in large-scale behavioral changes via tissue-specific physiological manipulation, often acting at very low doses or very slow time-scales. For example, queen fertility signaling is derived from the fertility signaling of the solitary ancestor [62], and honey bee forager flower-marking scents may serve related roles in solitary bees [24,63]. Even for conserved signaling pathways, radically different contexts between solitary and colony living result in almost unrecognizable enactment of the same functional means. For example, similar genetic and neurobiological pathways integrate nutritional state with foraging behavior in solitary flies and ants [29,64], but flies are fundamentally foraging to feed themselves, while ant foragers are activated to acquire nutrition for the colony.

Apart from hormones co-opted from solitary insect signaling pathways, eusocial insect regulatory network can integrate new players over evolutionary time, especially in novel tissues and in positions peripheral to gene regulatory networks [16]. These new players in gene regulatory networks can arise via duplication followed by neo-functionalization, or via origination of novel coding sequences from non-coding sequences. In either case, these taxonomically-restricted genes could play a crucial role in "sealing in" patterns of nestmate variation in physiology and behavior, for example by allowing task-specific evolution of coding sequences in a task-biased paralog pair, as seen in the case of insulin [65] and vitellogenin [66] signaling pathways. In mammals, it has been proposed that brain pathways can duplication and subspecialize, and thus elaborate over evolutionary time through processes similar to gene duplication [67]. It would be interesting to consider whether exocrine glands in social insects may also undergo duplication and subfunctionalization over evolutionary time, potentially facilitated by expansions in families of transcription factors and enzymes involved in the production of gland secretions.

Because of structural and algorithmic similarities, organismal physiology and eusocial colony physiology share common physical constraints and both have been modeled using similar approaches. The role of physiology, in organisms and colonies, is to maintain the system near functional attractors (homeostasis) and allow adaptive responses to environmental stimuli (learning & hormesis). These complex system-level properties must be maintained despite energetic demands and fundamental environmental uncertainty [54,68]. As a function of analogous ecological and functional constraints, similar system properties arise in organismal and colonial physiology. Overlapping topics and perspectives here include the use of models from information/communication theory [11] (signal-detection, threshold models, energetic constraints of bandwidth), decentralized decision-making [55,69], and evolutionary game theory [70,71]. The fundamental tradeoffs intrinsic to decentralized systems present themselves to both organisms and colonies: explore vs. exploit [72], modularity vs. specialization [11], signal vs. noise, performance vs. fragility [54], and adaptability vs. evolvability [73,74].

In many animal species, conspecific interactions influence physiology (e.g. pregnancy, group hunting/feeding, social status, etc. [75–77]). What is different in eusociality is that the colony is the evolutionary unit of behavior and physiology – colonies are selected to the extent that they leave more successful offspring, for example by producing more or better sexuals (reproductive males and females). In the obligately social eusocial insects, colony-level traits (e.g. architecture, efficacy of foraging) influence colony-level productivity (i.e. production of workers and reproductives) and thus colony ecological success [78,79]. Millions of years of selection for colony function results in radically reduced fertility of workers, and thus reduced costs of reproductive conflict (e.g. in ant species such as *Monomorium pharaonis* workers have no ovaries at all, in species like *Harpegnathous* workers can regain reproductive status but are unproductive in regular contexts). This extreme reproductive partitioning arises from the reoriented incentive structures in eusociality regarding honest vs. dishonest signaling [23]. Essentially, eusocial colonies beyond the "point of no return" are able to engage in runaway collaborative signaling (meaning the improved fidelity and efficiency of collaborative signaling systems) rather than semi-adversarial tit-for-tat signaling games [23,80]. The elaborate honey bee dance language for example, is a system of many signals coordinating the activities of two castes of bees based on completely honest communication. There are complex signaling games in other systems, in the context of sexual signaling for example, but in these cases a mixture of honest and dishonest signaling is present [81,82].

**Eusocial colony physiology: Hormonal Mechanisms and Evolutionary Consequences.**

In the eusocial insects, there is fundamental rewiring and turnover of the physiological mechanisms present in the solitary ancestors. After major transitions in evolution (prokaryote –> eukaryote & single cell –> multicellular life [41,83,84]) various structural differences exist in the derived (complexified) state relative to the ancestral pre-transition state. Look no further than the barely-recognizable mitochondria & its co-dependent master the nucleus. Insect species that are past the "point of no return" of being eusocial reflect the outcome of basic insect physiology (e.g. ancestral body plan, conserved gene families, physiological mechanisms), overlaid with a secondary phase reflecting the evolution of the physiological specialization found in caste polyethism & behavioral polyphenism. Even without any sort of morphological specialization, there can still be behavioral and pheromonal interactions between individual group members that can affect the physiology and development of each individual in the group [10]. Simply by joining together into a social group, additional possible regulatory mechanisms can be enacted at the collective level, for example bootstrapping task specialization off of pre-existing variation among conspecific individuals in body size [85] or behavior [86]. Signals that were used in a different context before group formation could be repurposed for new ends (e.g. sex pheromones used for male-female interactions could be used for intra-group interactions). A range of genomic and physiological novelties underlie the evolution of colony-level physiological processes (e.g. some are ancestrally mediated by pheromones, others may require the evolution of new paralogs). In this post-"Point of No Return" colony-level hormonal elaboration syndrome, we can consider several ways in which ancestral solitary insect physiological mechanisms (e.g. related to the regulation of individual development or foraging behavior) might have been shaped during the transition towards eusociality, and the functional genomic signatures of such pressures in the current day.

Several studies suggest that there is increased complexity of genomic regulatory mechanisms in eusocial insects (e.g., increased number of transcription factors or CREs [34,87,88]), and apparently a tightly constrained role for transposable elements [50,89,90]. Over evolutionary time, genes and signaling molecules can be gained or lost from regulatory networks. Gene regulatory networks can evolve via the addition of signaling hubs from other ancestral signaling networks through new connections (more common as per EvoDevo model [73]), resulting in novel phenotypes [91]. Alternatively, gene regulatory networks can grow by integrating novel (taxonomically-restricted) genes, facilitated by the fact that younger genes are apparently under less transcriptional coordination at the level of organs [16,92] and nestmate caste distinctions [93,94]. It appears that gene regulatory networks evolve through both changes in the regulation of conserved loci and incorporation of new players: novel loci are more likely to be incorporated into distal parts of gene regulatory networks and be expressed in novel or secretory tissues, while conserved loci are more likely to undergo changes to transcriptional regulation in conserved tissues [16,95].

The function of conserved members of physiological regulatory processes can be influenced by sequence changes, new regulatory connections, or other contextual changes. For example the cGMP-dependent protein kinase G enzyme (known as *foraging* in *Drosophila*) is famous for being a conserved player in the neurobiology of foraging and metabolic regulation across vertebrate and invertebrate taxa [96–98]. While the homology of the PKG locus is indeed deeply conserved, the action of PKG is cell-type specific and also probably depends on the identity of downstream phosphorylation targets. Hence there is not a consistent role or direction of effect for PKG even across just Hymenoptera [99–104]. Thus while PKG may play a role in foraging-related physiological networks of diverse insects, there are species-specific changes to the inputs & outputs of PKG such that the function of over- or under-expression of PKG cannot be reliably predicted, even locally. Similar claims could be made for "conserved" gene families such as pigmentation/neurotransmitter-related genes that now play roles in regulating worker behavior [30,105], and conserved neuropeptides that have gained task-specific functions [32,65]. Recent evidence suggests that genetic pathways involved in generating sexually dimorphic morphology and behavior in solitary insects (e.g. dsx/fru/tra [106–108]), are involved in the caste differentiation in eusocial insects [109,110]. This suggests that the gene regulatory networks that orchestrate variation among the physical castes in the eusocial insect colony may be as extensive as those underlying sexual dimorphism in solitary insects, potentially even reusing many of the same molecular components [111].

**Some hormonal processes that are mediated internally in other insects, may fall under the control of another task group within obligately eusocial colonies.** While the basic players of a physiological mechanism may remain the same, there may be a spatial reorganization of signaling so that regulation is enacted across multiple insect bodies. For example, foraging behavior and fat metabolism are linked through integrated neurohormonal mechanisms in *Drosophila* [64], such that flies forage when hungry and stop feeding when full. Eusocial insect colonies must also balance foraging behavior with fat metabolism and food reserves, with an additional challenge: the foraging behavior is performed by an entirely disjoint set of nestmates (foragers) from those engaged in fat metabolism (larvae). These behaviorally- and physiologically-specialized components of the colony engage in cross-regulation using behavioral interactions [112–114] and chemical signaling [115], and when these regulatory feedback systems are pushed beyond their limits, colony collapse is the result [116]. The exact mechanics of the physiological decentralization in the eusocial insect colony will depend on species-specific colony structure and life history. For example stingless bees seal larvae into cells with provisions, while ants feed brood continuously through the larval instars.

The Reproductive Groundplan Hypothesis [12,65,117] (& other Toolkit-like hypotheses [118]) posits that the seasonally-oscillating ecological demands lead to phenotypic plasticity of the ancestral Ur-ant (between forager-like and queen-like states). This evolutionary history is reflected today by the partitioned expression of the same genome between workers and queens [29,119,120]. This plastic state is still observed in species considered to be "facultatively social" (note that this label does not imply that all social species progress along similar or predictable evolutionary paths [18]). However in obligately eusocial insects, millions of years of evolution have shaped colony function such that tissue- and caste-specific specialization no longer exists within the bounds of any plausible ancestral phenotypic plasticity. In extant eusocial taxa we observe behavioral and physiological extremes that are far beyond the range of any solitary species (e.g. 30+ year queen life in *Pogonomyrmex*, agriculture & extreme polyphenism of *Atta*, developmental scaling of *Pheidole*, workers without ovaries in *Monomorium* and *Brachyponera* [121], etc.). These extreme derived states are facilitated and accompanied by significant alterations to the hormonal pathways involved in generating these phenotypes relative to the proto-eusocial ancestor or contemporary solitary insects.

**Case Studies in colony physiology: Ancestral traits under colony control, and Novel colony traits.**

There are broadly two kinds of phenotypes (measurable traits or characteristics) of eusocial insect colonies. First there are phenotypes that can be measured from a single individual insect such as head width or ovary number – these traits can be modeled within the framework of a worker being an individual that receives input from other conspecifics). Second, there are traits that are the outcomes of collective behavior and as such cannot be reduced to physiology of nestmates, for example nest architecture or rate of brood production. Traits of the first kind, which manifest as variation in nestmate morphology or gene expression, bear direct homology to traits of solitary insects. In eusocial insects, these bodily traits have fallen under extensive control of other nestmates via social physiology [12,48,49]. The second kind of traits are not simply modifications of insect body physiology, as they reflect colony-specific adaptations to colony living. These truly collective traits arise from the interaction of nestmates and the environment, and unsurprisingly the mechanisms that regulate these colony traits are unconnected or functionally absent in solitary insects. Here we cover several case studies that reflect the broad range of physiological elaborations we see in eusocial insects, drawing on examples of traits that manifest at the insect or colony levels

**Colony physiological regulation of body traits of nestmates: under altered physiological control and novel colony-level inputs**.

Regulation of female fertility and reproduction is the crux of the eusocial colony lifestyle. Within a eusocial insect colony, the reproductive skew between queens (who can lay thousands of eggs) and workers (who usually lay zero eggs, and often are lacking ovaries entirely) is extreme. These differences in fertility are linked to morphological, hormonal, and transcriptomic differences in essentially every tissue of the body. Regulation of fertility in the eusocial insect colony occurs via a variety of mechanisms that are all essentially absent in solitary insects. These mechanisms include short-range chemical signaling [122], control of nutritional intake [123], and multiple modes of physical interaction such as piping and drumming [124]. In various ant and bee species, secretions passed among workers and queens can influence the fertility of all engaged actors, and thus influence colony productivity overall. In pharaoh ants and fire ants, there is good evidence that queen fecundity is strongly affected by the presence of larvae, as well as the anal and oral secretions made by larvae of specific stages [125,126]. Honey bee queens are stimulated to produce more eggs by being exposed to brood pheromone [127], a positive feedback cycle within the colony where egg-laying stimulates more egg-laying. Another primary semiochemical regulator of fertility in honey bees is queen mandibular pheromone (QMP). QMP is produced by active queens and has the effect of suppressing fertility and inducing other physiological changes in nearby workers, thus it is a negative feedback signal. The downstream targets of queen pheromones are partially conserved among *Lasius* ants and *Apis* and *Bombus* bees [128] despite vast evolutionary and ecological differences among these species. Additionally, honey bee QMP has phenotypic effects on *Drosophila* adults in the same direction as bees (e.g. repression of fertility in females), and also triggers behavioral changes in males [129]. This is consistent with the notion that colony physiological decentralization may arise and be stabilized through the reuse of pheromonal mechanisms that are present in solitary insects, acting through conserved or novel inputs and outputs.

Worker physiology is shaped by colony context in eusocial insects, through the use of many feedback loops and signaling systems. Here we focus on how several central hormones are involved in coordinating task-specific behavior through conserved and derived regulatory connections. **Corazonin** is a pleiotropic invertebrate hormone that is orthologous to the vertebrate Gonadotropin-releasing hormone (GnRH) [53,130,131]. In *Drosophila* fruit flies, Corazonin plays a role in coordinating metabolism to deal with stressful states [132]. In flies and other solitary invertebrates, corazonin plays a role in regulating both male and female reproductive biology [133–135]. In the social insects, corazonin is expressed in workers, especially foraging individuals [32]. In *Harpegnathos* ants, worker levels of corazonin are correlated with their foraging activity, and injection of corazonin suppresses the expression of vitellogenin and influences behavior [32].

**Vitellogenin**-family genes (Vg) are egg yolk proteins in solitary insects, and may also have behavioral roles in the brain [136]. Vg-related proteins belong to an ancient gene family that is diverse in both vertebrates [137] and invertebrates. Several factors complicate the analysis of the roles of Vg in the ants and bees. First, there are multiple related Vg paralogs within each eusocial insect species, often with caste- or age-biased expression [66,138]. This suggests that as the Vg gene family content changed over evolutionary time, ancestral functions of Vg are altered and partitioned into multiple contemporary Vg-family proteins. Second, Vg is known to play tissue-specific roles in various essential body processes, for example relating to development, immunity, and inflammation [139,140]. Thus it may be difficult to disentangle the specific behavioral role of Vg, or other highly pleiotropic players, in eusocial insects. A key hormone linked to Vg signaling is **Juvenile Hormone (JH)**, an ancient sesquiterpenoid regulator of growth and differentiation in invertebrates. JH is transferred between nurses and larvae during trophallaxis, along with other bioactive compounds [27]. Thus the JH-regulated systems, of which the major components exist in solitary insects, have fallen under multi-body regulation in the eusocial insects.

**Predictions for Eusocial Colony Physiology.**

Here we present predictions regarding evolutionary & functional genomics of behavior and hormones in eusocial insects, summarized in Table 1. These predictions set a course for the integrated understanding of colony function as arising from nestmate specialization and decentralized physiological processes that have been shaped by millions of years of colony-level selection. All predictions and hypotheses should be carried out within a phylogenetic comparative framework to disentangle the relative importance of genomic, ecological, and behavioral constraints over evolutionary time [1,30,141,142].

**Table 1**. Predictions arising from a decentralized perspective on eusocial colony physiology.

| Area of Prediction | Hypotheses |
|---|---|
| Mechanisms of physiological regulation | ● Increased role of colony-level processes in the regulation of nestmate behavior, gene expression, and hormonal state.<br>● Increased complexity of physiological regulatory connections and altered input/outputs of conserved genes<br>● Feedback loops across multiple timescales will involve multiple distinct mechanisms derived from solitary insects (nutrient transfers, mechanical stimulation, stigmergic pheromone deposition & changes to nest architecture). |
| Game Theory | ● Increased fidelity and decreased negative pleiotropy of interactions among eusocial nestmates relative to social interactions among non-eusocial nestmates<br>● Decreased role of individual decision-making as a consequence of offloading to collective processes. |
| Glands | ● Increased number and type of glands in eusocial species.<br>● Caste- and task-specific functions for both conserved and novel glands, especially involving secretions of (metabolic products of) taxonomically restricted genes. |
| Development | ● Increased importance of interactions among developmental stages in regulating nestmate behavior.<br>● Bidirectional transfer of developmental modulators among individual larva and nestmates. |
| Gene regulatory networks and Signaling pathways | ● Regulatory networks of solitary ancestors will be decentralized across multiple nestmates.<br>● Caste- and task-specific utilization of gene regulatory networks |

| | will be in feedback with physiological specialization of nestmates. |
| | ● Non-linear changes in colony traits due to caste-, tissue-, and task-specific changes in gene expression. |

**Predictions for glands**: We predict that eusocial insect species will have nestmates with more exocrine glands (e.g. duplication and neofunctionalization of conserved glands), and glands with more complex or voluminous glandular secretion when compared with solitary insects. Further, there may be patterns within eusocial taxa such that species with higher social complexity may have more specialized glandular structure present across nestmates. From an evolutionary signaling theory perspective, once a nestmate exocrine gland has become fully coopted into colony-level regulatory networks, its dynamics and constraints will approximate that of organismal endocrine glands. Thus we hypothesize that chemical stimuli shared among nestmates have been selected for high-fidelity and rapid coordination of colony physiology to changing demands. We predict that eusociality provides a new context for signaling systems such that the nature of the chemical signaling among nestmates can be more elaborate than social cues in non-eusocial species [23]. We predict that the transfer of direct mediators of insect physiology among nestmates (microRNAs or chromatin remodelers in *Apis* royal jelly [143], hormones in nurse feeding fluid [27]) will not induce antagonistic responses observed in solitary insects (such as sex conflict in *Drosophila* [144]), even when some of the same molecules may be used. Within the colony, honest signaling among nestmates can flourish, it is not an arms race as with conspecific interactions in social species or symbiotic relationships among different taxa. To that extent that "cheating" or "free-riding" behavior exists within eusocial insect colonies, we consider such phenomena to be exceptions that prove the rule, not the foundation from which eusocial colony function rests upon.

**Predictions for signaling pathways**: We predict that elaboration & partitioning of ancestral signals will occur such that receptors, signaling pathways, and metabolic pathways that are expressed by the solitary ancestor over the course of a lifespan are expressed synchronously but distinctly by various physiological castes of the colony. The exocrine glands & chemosensory organs [36,145] in particular can be expected to have strongly partitioned expression among castes and tasks [36,145]. One challenge for transcriptomic and epigenomic studies of brain function is that the brain undergoes many types of physiological changes for which gene expression changes are delayed, complex, or absent (e.g. topological changes in neural circuits, protein modification at synapses, time lags between neural transcription and translation). This can be contrasted with exocrine glands, for which the transcriptome can be expected to more closely approximate the instantaneous secretory function of the tissue due to rapid transcriptomic turnover [16,95]. We expect that integrated neuroscientific approaches involving the single-cell profiling of social insect brain tissue along with live-imaging and reverse genetic approaches will be required to reach nuanced claims about the neurophysiology of individual behavior [88,146]. However it is a second layer of organization above individual neurobiological mechanisms, through interactions among nestmates, by which colony collective behavior arises [54,55,69]. Consistent with this decentralization of cognition across multiple nestmate bodies, colonies with increased size and specialization may have workers with proportionally smaller brains [147,148]. Another implication of increased physiological specialization is that genes with task- and tissue-specific expression patterns may be associated with non-linear changes in colony collective behavior, for example by altering worker sensitivity to interactions or ambient conditions [149–151].

**Predictions for gene regulatory networks**: Gene regulatory networks of derived organismal colonies may be more complex than those controlling solitary insect physiology and behavior [12,48,152]. Here we mean that eusocial regulatory networks are more complex in the sense that they allow for a broader range of functional connections among genes (through interactions among nestmates), increased spatial partitioning of expression (e.g. novel sex-, caste-, and tissue-specific expression patterns), and novel expression patterns through developmental time (e.g. polyethism). Additionally these derived eusocial regulatory networks can be considered more complex in that they allow the colony to exhibit emergent behaviors that do not exist in solitary insects, such as nest architecture and brood care. This hypothesis that there is increased regulatory complexity in the eusocial insects is empirically confirmed by studies showing that eusocial species have unique patterns of transcription factors, cis-regulatory elements, and epigenetic regulatory mechanisms [34,92,115], potentially reflecting the demands of this new eusocial mode of colony physiological regulatory mechanisms. This also points to the advantages of taking a network approach to analyzing gene expression data, in addition to traditional differential expression statistics [93,150,153]. Key general questions here include how ecological factors interact with solitary insect gene regulatory networks in order to allow for the transition to eusociality [18,154], and in which ways these transitions toward eusociality are unique vs. universal [18,71,155].

We predict that novel gene regulatory networks will be formed from this decoupling of otherwise conserved pathways and traits. This is because the colony context allows for regulatory links to arise among nestmates in different developmental stages (e.g. signaling between larvae to adults [152]), as well as utilizing physiological regulatory connections involving tactile and vibratory mechanisms [156,157]. This means that there is the potential for diversified types of signaling and response in the eusocial insect colony, as well as elaboration of the molecular mechanisms underlying the response to stimuli. Functional genomic approaches that simultaneously consider multiple socially-interacting individuals, e.g., based on sequencing a time series of interacting nurse workers and larvae [152], can begin to disentangle the molecular mechanisms of social signaling and the downstream physiological and developmental response. Exocrine gland and endocrine glands that are linked up within the same physiological pathway in a colony (e.g. regulating foraging or reproduction) are never in the same network in solitary insects, and this derived state should be explicitly considered when performing pathway analysis or using other functional genomic approaches. Further, work on signaling pathways related to JH and Vitellogenin and shows that even the most fundamentally important conserved genes have distinctly different expression patterns in eusocial insects as compared to solitary insects, as well as expression variation between related eusocial species and among nestmates [31,158–160]. The convention has been to act as if use of an ortholog constitutes conservation, but already for key cases such as PKG we know the same locus can be associated with a trait (e.g. "playing a conserved role") yet still have unpredictable patterns of expression or functional roles. Holistic (i.e. colony-level) consideration of these issues is necessary to understand how selection acts to shape gene regulatory networks that play out across multiple insect bodies [48]. For example, a recent study in honey bees found that decades of artificial selection for increased royal jelly production was accommodated by changes in the expression of chemoreceptor proteins in nurse antenna [151]. This can be understood from the perspective that nurse antennae are one of the multiple tissues that are involved in the emergent regulation of colony reproductive investment and royal jelly production. In other words, colonies may respond to evolutionary and ecological challenges in a non-linear fashion, via shaping the expression of genes that influence tissue-specific physiology of sensory organs and brain signaling processing [161,162].

**Future directions & questions:**

There are many opportunities for functional genomics to use eusocial insects as model systems to address general questions about hormones, development, and behavior. First, the epigenetic plasticity of eusocial insect workers situates them as tractable models to disentangle genetic and environmental influences on behavior [88,163]. The ecological diversity of the eusocial insects provides broad possible scope for understanding how colonies solve niche-specific challenges, and since many species can be kept in the laboratory so that genetic and environmental factors can be controlled. Second, new techniques can be integrated in eusocial insect taxa to bring about multidisciplinary synthesis. Recent and ongoing studies are combining natural history, automated behavioral analysis, DNA/RNA-sequencing, transgenic techniques, and pharmacological manipulations [29,120,146,155,164].

A key question is: How dramatic are the molecular changes necessary for the major transition to eusocial colonial living from a solitary or social state? Several previous authors have stressed that few molecular changes may be necessary for the initial transitions from solitary insects to small eusocial colonies [165,166]. We emphasize that in lineages with large and complex eusocial colonies, extreme molecular changes have likely occurred that obscure the traces of initial molecular inroads towards colony living [94,167]. Thus it is important to consider to what extent the multiple independent origins of eusociality used convergent versus taxa-specific mechanisms [18]. New tools allow us to do many things in non-model systems that previously could only be done in model systems, and systems like *Drosophila* have proven helpful in broad strokes for elucidating insect physiological mechanisms. However millions of years of selection for colony function in eusocial insects means that even for conserved orthologs (e.g. PKG, biogenic amine receptors), functional gene roles may differ. This is a significant issue for Gene Ontology (GO) based analysis of functional genomic experiments in eusocial insects, as most GO terms in these species are directly transferred from *Drosophila*. Any analysis of eusocial insects that is templated off of a (distantly related) solitary insect species will systematically ignore the role of taxonomically restricted genes [94], overstate the role of orthologous genes, and unable to consider the implications of decentralized colony physiological mechanisms. The challenges of colony living in eusocial insects have been accommodated through multiple types of genomic and epigenomic changes, and research should highlight these taxa-specific adaptations, not average over them. If the goal is to gain unbiased insight into the genetic changes that are most biologically important – as opposed to the set of genetic changes that involve highly conserved genes with more-or-less well-characterized functions in solitary organisms – then alternate approaches may be required. For example, maybe we need to put the model system approach on the back burner and start to look at genes that seem biologically important in eusocial insects, independent of whether they are found in other insects.

Promising experimental approaches in the social insects could use RNA-Seq, proteomics, and metabolomics on the same tissue-specific samples across the classes of individuals (e.g., developing larvae, adult nurses [152]) involved in colony physiological processes. It is especially interesting to combine these functional genomic analyses with computational methods such as the automated tracking of behavior from video data [114,155,168]. For example worker-level tracking can assess how worker heterogeneity leads to colony foraging performance [169,170], or how trophallaxis networks provide robustness to variability in colony resource intake [171]. Specifically these types of studies in eusocial insect species could connect multilevel-network perspectives on animal behavior [11,172] with the molecular mechanisms of behavioral epigenetics and neurophysiology [29,119,162], in the context of a group of species with diverse ecologies and rich natural history.

**Animal Welfare statement:** No animals were used in our research for this paper.

**Declaration of interest:** None.

**Funding**: TAL was funded by National Science Foundation grant IOS-1452520.